\documentclass[aps,pra,reprint, superscriptaddress]{revtex4-1}

\usepackage{epsfig,graphics,graphicx}% Include figure files
\usepackage{dcolumn}% Align table columns on decimal point
\usepackage{bm}% bold math
\usepackage{color}
\usepackage{amsmath,amssymb}
\usepackage{fullpage}
\usepackage{units}
\usepackage[usenames,dvipsnames]{xcolor}
\usepackage[utf8]{inputenc}
\usepackage[T1]{fontenc}

\usepackage{mathpazo}
\usepackage{dsfont}

\usepackage{mathptmx}
\usepackage{amssymb}
\usepackage{amsmath}
\usepackage{latexsym}
\usepackage{amsfonts}

\usepackage{hyperref}
\hypersetup{pdfpagemode=UseNone}

\newtheorem{lemma}{Lemma}
\newtheorem{theorem}{Theorem}
\newtheorem{proposition}{Proposition}
\newtheorem{definition}{Definition}

\newcounter{rem}
\newcommand{\bra}[1]{\ensuremath{\left\langle#1\right|}}
\newcommand{\ket}[1]{\ensuremath{\left|#1\right\rangle}}

\def\>{\rangle}
\def\<{\langle}

\renewcommand{\rho}{\varrho}
\newcommand{\idty}{\mathbb{1}}
\def\tr{{\rm Tr}}

\def\ii{{\rm i}}

\def\textbf#1{{\bf #1}}

\def\beq{\begin{equation}}
\def\eeq{\end{equation}}
\def\be{\begin{equation}}
\def\ee{\end{equation}}
\def\ben{\begin{eqnarray}}
\def\een{\end{eqnarray}}
\def\beqa{\begin{eqnarray}}
\def\eeqa{\end{eqnarray}}
\def\eea{\end{array}}
\def\bea{\begin{array}}
\newcommand{\bei}{\begin{itemize}}
\newcommand{\eei}{\end{itemize}}
\newcommand{\bee}{\begin{enumerate}}
\newcommand{\eee}{\end{enumerate}}
\def\bep{\begin{proposition}}
\def\eep{\end{proposition}}
\def\bel{\begin{lemma}}
\def\eel{\end{lemma}}
\def\bet{\begin{theorem}}
\def\eet{\end{theorem}}
\def\bed{\begin{definition}}
\def\eed{\end{definition}}
\usepackage{soul}

\begin{document}

\title{Reliability of digitized quantum annealing and the decay of entanglement}

\author{John P. S. Peterson}
\affiliation{Centro Brasileiro de Pesquisas F\'{i}sicas, Rua Dr. Xavier Sigaud 150, 22290-180 Rio de Janeiro, Brazil}

\author{Roberto S. Sarthour}
\affiliation{Centro Brasileiro de Pesquisas F\'{i}sicas, Rua Dr. Xavier Sigaud 150, 22290-180 Rio de Janeiro, Brazil}

\author{Alexandre M. Souza}
\affiliation{Centro Brasileiro de Pesquisas F\'{i}sicas, Rua Dr. Xavier Sigaud 150, 22290-180 Rio de Janeiro, Brazil}

\author{Ivan S. Oliveira}
\affiliation{Centro Brasileiro de Pesquisas F\'{i}sicas, Rua Dr. Xavier Sigaud 150, 22290-180 Rio de Janeiro, Brazil}

\author{Frederico Brito}
\email[]{fbb@ifsc.usp.br}
\affiliation{Instituto de F\'{i}sica de S\~ao Carlos, Universidade de S\~ao Paulo, C.P. 369, S\~ao Carlos, SP, 13560-970, Brazil}

\author{Fernando de Melo}
\email[]{fmelo@cbpf.br}
\affiliation{Centro Brasileiro de Pesquisas F\'{i}sicas, Rua Dr. Xavier Sigaud 150, 22290-180 Rio de Janeiro, Brazil}

\date{\today}

\begin{abstract}
We performed a banged-digital-analog simulation of a quantum annealing protocol in a two-qubit Nuclear Magnetic Resonance (NMR) quantum computer. Our experimental simulation employed up to 235 Trotter steps, with more than 2000 gates (pulses), and we  obtained a protocol success above 80\%. Given the exquisite control of the NMR quantum computer, we performed the simulation with different noise levels.  We thus analyzed the reliability of the quantum annealing process, and related it to the level of entanglement produced during the protocol. Although the presence of entanglement is not a sufficient signature for a better-than-classical simulation, the level of entanglement achieved relates to the fidelity of the protocol.
\end{abstract}

%\pacs{}
\maketitle

\section{Introduction} Among the models for quantum computation, quantum annealing arises as one of the front runners that may first establish the quantum supremacy -- the stage at which implementations of quantum computers will start solving problems deemed intractable for their classical counterparts. Such a model is inspired on the adiabatic quantum computation (AQC) scheme, originally proposed by Farhi et al.~\cite{Farhi20042001}, in which the answer of an abstract problem can be encoded in the ground state of a physical system. Similarly to AQC, quantum annealing exploits the gradual modification of the system state character, in order to find the solution of a hard problem starting from an easy one. However, regarding its physical implementation, quantum annealing presents a key advantage, since it is tailored for scenarios in which the system is in contact with a thermal environment -- due to this interaction, under appropriated conditions, relaxation processes to the ground state may enhance the protocol success. Because of that, and due to its simplicity, quantum annealing has attracted great attention. For instance, it was adopted as the quantum computation model by D-Wave~\cite{Johnson:2011gd} -- the first company commercially producing and selling devices advertised as quantum computers. This first private venture was recently followed by an initiative from Google/UCSB~\cite{Barends2016}.

All that led to an increased scrutiny of the model. Indeed, soon after the announcement of D-Wave's first machine, took place an important and intense debate whether their computer would be actually a quantum computer. Even though the first evidences indicated that quantum annealing would be the right model for the machine's behavior~\cite{Johnson:2011gd,Boixo:1ha,Dickson:1bv, Boixo:2014cg, PhysRevA.91.042314}, they were taken as disputable, as semi-classical approaches could also reasonably describe the experimental results~\cite{10.3389/fphy.2014.00052, 2014arXiv1401.7087S,2014arXiv1404.6499S}, and no evidence of speed-up was found~\cite{Ronnow:2014fd}. In such a debate, naturally, the ``holy grail'' became whether the machine could generate entanglement during the computation. Recently D-Wave conducted an experiment that unequivocally showed the presence of entanglement among the qubits composing one of their first processors~\cite{PhysRevX.4.021041}. Having settled that issue, other questions became natural and pertinent.

The aim of this contribution is twofold: \emph{i) To assess the reliability of a quantum annealing simulation under a massive ``banged-digital-analog'' quantum computation.} Recently, a only-digital simulation of the quantum annealing process was performed~\cite{Barends2016} in a system composed of nine superconducting qubits. Due to their system size and the noise acting on it, they were able to perform only few (five) Trotter steps, and, for the ferromagnetic chain problem with 4 spins, the fidelity obtained was $0.55$. To overcome some of the issues encountered in this implementation, here we combine the digital simulation with an analog part. Digital-analog quantum simulations have been proposed to different architectures~\cite{Arrazola,Lamata}. Such a scheme might potentially lead quantum annealing to inherit features, like designing interactions on demand and error correction protocols, from  digital quantum computing~\cite{Lloyd1073}.  \emph{ii) To relate the amount of entanglement generated during a quantum annealing protocol with its success.} The role of entanglement is one of the most unclear questions concerning AQC. Our results suggest that, once fixed an annealing schedule, its fidelity shall be related to the amount of entanglement created during the protocol. Therefore, only high levels of entanglement, i.e., extreme experimental control of noise sources, might guarantee a better-than-classical result. In order to tackle these issues, here we employed a two-qubit NMR quantum processor to perform the banged-digital-analog quantum computation. In NMR the interaction among the nuclear spins, which are the qubits of the processor, is always on, performing the analog part of the computation. Besides that, one can shine the system with radio frequency (rf-) pulses to perform banged-digital single qubit gates. Moreover, the amount of entanglement generated in the protocol can be changed by a tunable source of decoherence.

\section{The computational problem} The task we analyse here is that of finding the ground state of an Ising spin glass model, defined by the Hamiltonian:
\begin{equation}
H_{{\rm Ising}}=\sum_{i=1}^N h_i\sigma_i^z+\sum_{i<j=1}^N J_{ij}\sigma_i^z\sigma_j^z.\label{ising}
\end{equation}
In this expression, $\sigma_i^z$ is the usual $z$ component of a spin-1/2 operator at site $i$, and the parameters $h_i$ and $J_{ij}$ represent local fields and spin-spin couplings, respectively.
Besides being a paradigm for many-body quantum systems, the problem of finding its ground state is known to be representative of several  optimization problems (NP-hard)~\cite{Barahona:1982gj}. Note that if we define the eigenvectors of $\sigma^z_i$ by $\sigma^z_i\ket{k_i}= (-1)^{k_i}\ket{k_i}$ with $k_i\in\{0,1\}$, then  the ground state of the Ising Hamiltonian is certainly  a product state of the form $\bigotimes_{i=1}^N \ket{k_i}$. The challenge is  to determine which of the $2^N$ possible product states of this form is the actual ground state. A task for which brute force search clearly will not be efficient.

The quantum annealing strategy to approach this problem relies on the quantum adiabatic theorem~\cite{messiah}. First, one initializes the system in the ground state of a simple Hamiltonian, $H_\text{easy}$. Here we choose $H_\text{easy}=\sum_i \Delta_i \sigma_i^x$, with $\Delta_i>0~\forall i$, which ground state is $\bigotimes_{i=1}^N \ket{-_i}$, with $\ket{-_i}:= (\ket{0_i}-\ket{1_i})/\sqrt{2}$. Thus, the initial state is an equiprobable superposition of all $2^N$ possible product states $\bigotimes_{i=1}^N \ket{k_i}$ that can be the ground state of $H_\text{Ising}$.
Second, a ``schedule'' is chosen such that the system Hamiltonian is adiabatically changed onto the Ising Hamiltonian. Specifically, the system is governed by the time-dependent Hamiltonian: 
\beq
H(t)=\Gamma(t)H_{{\rm{easy}}}+\Lambda(t) \sum_{i=1}^N h_i\sigma_i^z + \Omega(t)\sum_{i<j=1}^NJ_{ij}\sigma_i^z\sigma_j^z,
\label{eq:timeH}
\eeq
where the envelope functions $\Gamma$, $\Lambda$, and  $\Omega$ are changed smoothly during the protocol, $t\in [0,T]$, and are such that  $\Gamma(0)\gg\Lambda(0)\approx \Omega(0)$ and $\Gamma(T)\ll\Lambda(T)=\Omega(T)$, what ensures the appropriate initial and final conditions. If these conditions are met, then  the quantum adiabatic theorem guarantees that the final state of the system is the ground state of $H_\text{Ising}$. It is worth noticing  that, as for  $t>0$ the spins are interacting and $[H(t),H_\text{easy}]\neq0$, thus it is expected that some entanglement will be generated during this process.

Naturally, any physical implementation of the quantum annealing protocol must run in a finite time, and is under the influence of a thermal environment. The finitude of the protocol duration implies that the adiabaticity condition is somewhat broken, and the state at each time is a superposition of a large component of the ground state and small parts of the first excited states. Furthermore, if the protocol is slow enough and the system and the environment are weakly coupled, the thermal environment turns this superposition into a mixture of such eigenstates. These two unavoidable facts, in general, result in errors for the quantum annealer. However, since we are searching for the ground state of the problem Hamiltonian, for low temperatures the thermalization process might help transferring population from the excited states to the ground state. This possible improvement of the quantum annealer comes at the cost of keeping the temperature low, and of waiting for the relaxation process to happen.

In the following, we  address the robustness of the quantum annealer in such realistic conditions -- fixing the running time for different noise strengths -- and we compare the quality of its results with the ones obtained by a classical simulation. In the classical simulation here employed, we model each spin as a magnet compass,  with magnetization pointing at the direction $\vec{M}_i= (M^x_i,M^y_i,M^z_i)$. The magnetization components of each qubit evolve accordingly to the noise-free Bloch's equations:
\beq
\small{
\frac{d\vec{M}_i}{dt} = \left(-\Gamma(t)\Delta_i \hat{e}_x - \Lambda(t) h_i \hat{e}_z -\Omega(t)\sum_j J_{ij}M_j^z \hat{e}_z\right)\times \vec{M}_i.}
\eeq
This set of $3N$ coupled differential equations can be easily solved by a classical computer. We  say that the state of the system at time $t$ is given by $\bigotimes_i^N\ket{\psi_i(t)}$, with the state of each qubit defined by magnetization components through $\bra{\psi_i(t)} \sigma^j_i \ket{\psi_i(t)}= M_i^j(t)$ for $j\in \{x,y,z\}$.
In this classical model  there is clearly no entanglement during the whole evolution. As such we expect that this simulation will fail during the time intervals where the quantum annealing process produces some entanglement. As a last remark, since this is a simulation performed in a classical computer, we do not include any noise effects. The state assigned to the system remains always pure.

\section{NMR  Experiment} In order to address the reliability of a quantum annealing process in controlled laboratory conditions, we conducted an experiment of a small-scale quantum annealer within the framework of nuclear-magnetic resonance (NMR). NMR is a well established test bed for quantum information processing, allowing for an exquisite control of the nuclear spins of molecules. Indeed, one of the first implementations of an adiabatic quantum optimization algorithm already reported was one using an NMR experiment\cite{matthias}.

Our experiment was performed using a sample of carbon-enriched Chloroform (CHCl$_3$) molecules with C$^{13}$, where the nuclear spins of the Hydrogen and the Carbon, both having spin 1/2, were taken as physical implementations of qubits.  For that, a static magnetic field $\mathbf{B}_0$ of approximately $11.74\rm{T}$ was applied to the sample along the $z$ direction, yielding $\omega_H/2\pi = 500\rm{MHz}$ and $\omega_C/2\pi = 125\rm{MHz}$ as the Hydrogen and Carbon Larmor frequencies, respectively. The sample contains around $10^{16}$ identical Chloroform molecules highly diluted in $97\%$ of deutered Acetone, thus intermolecular interactions can be safely ignored. Such conditions lead to the natural NMR system Hamiltonian:
\beq
H_{\rm NMR} = -\hbar\omega_H \frac{\sigma^z_H}{2}-\hbar\omega_C \frac{\sigma^z_C}{2} +2\pi\hbar J \frac{\sigma^z_H \sigma^z_C}{4},
\label{eq:Hnmr}
\eeq
with the interaction strength between the two spins given by the coupling constant $J=215{\rm Hz}$.

In order to perform the operations required, a Varian 500 MHz NMR spectrometer was used, which has two channels allowing for the manipulation of two nuclear spins simultaneously. We resort to rotating frames of reference, such that the effect of local Hamiltonians (due to static field and spurious frequency displacements) are canceled, and focus can be set on the interaction part. Single qubit operations are performed in a straightforward manner by applying radio frequency pulses with specific field polarizations, tuned in resonance with each one of the spins (see Appendix A). However, since pulses and read-out act collectively, it is not possible to address each molecule individually, meaning that only average properties of the sample are measured~\cite{NMRreview,IvanLivro}.  Another consequence of such a lack of spatial resolution is that a magnetic field gradient along the $z$ direction acts as a dephasing channel for the computational basis (see Appendix A). Given that the strength of this process is determined by the gradient intensity, one has an effective \emph{knob} to adjust the noise level in this setup. Furthermore, as the system's Hamiltonian is not diagonal in that basis, eigenenergy transitions may also be induced by the presence of such  magnetic field. We exploit this tool to assess the connection between different levels of entanglement and the fidelity of the noisy computation.

\begin{figure}[t!]
\begin{center}
\includegraphics[width=\linewidth]{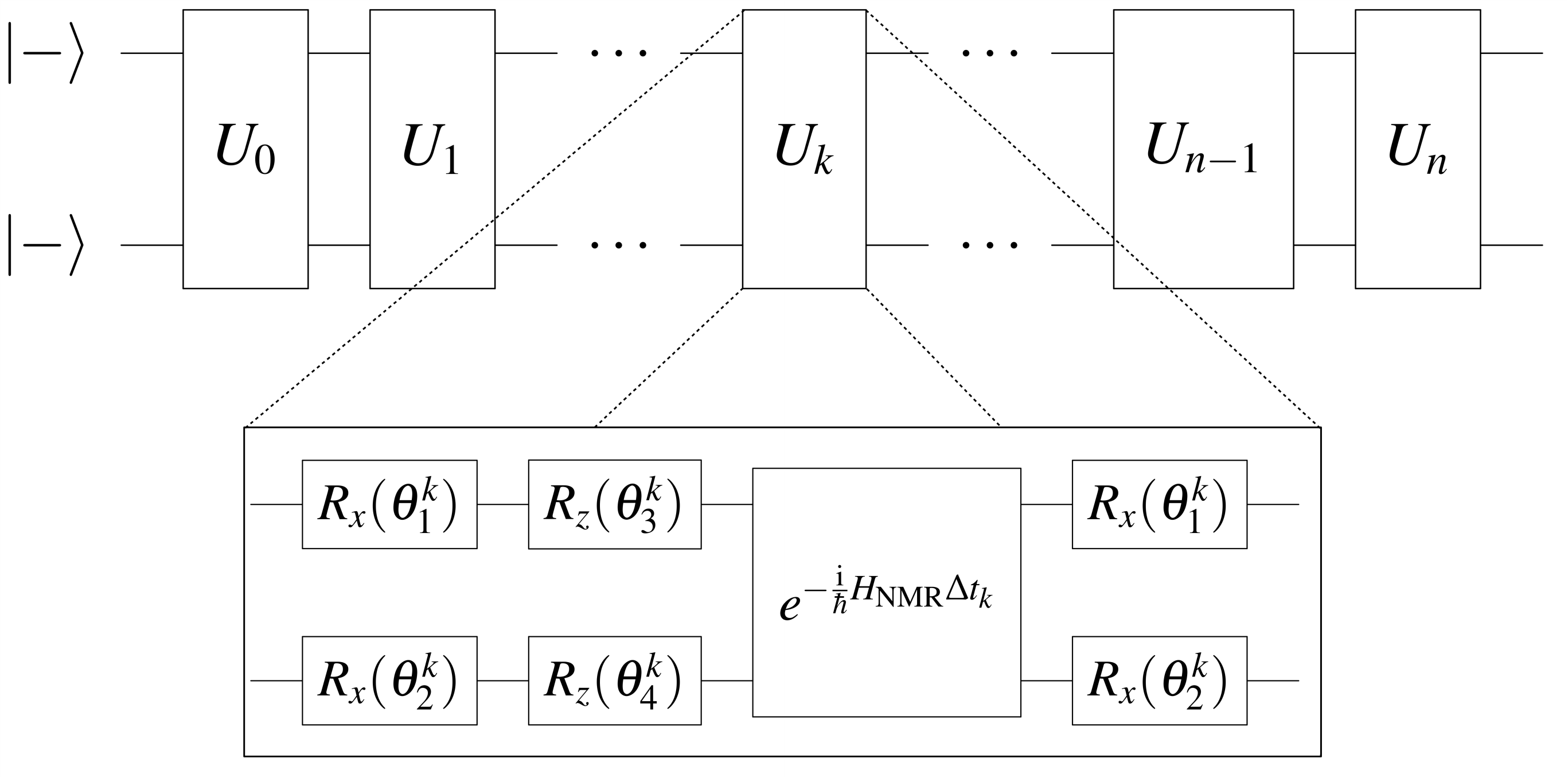}
\caption{\footnotesize \textbf{Simulation of the quantum annealing protocol.} The total time evolution is constructed as a sequence of evolutions $U_k:=U(t_{k+1},t_k)$. Each evolution block $U_k$ is implemented using free evolution and an appropriate  $z$ and $x$ polarized rf-pulses applied to each qubit. The respective pulses angles $\theta_i^k$ and free evolution duration $\Delta t_k$ are chosen as to maximize the fidelity of the operation with $U_k$.  \label{fig:simulation}}
\end{center}
\end{figure}

Our quantum annealer is then comprised of two qubits and its time evolution is simulated experimentally using the natural (analog) NMR Hamiltonian~\eqref{eq:Hnmr} and an appropriate (digital) sequence of pulses. For that, we first divided (Trotterization) the total evolution generated by the time-dependent Hamiltonian~\eqref{eq:timeH} into 235 time steps $\delta t_k$, with $k\in \{0,\ldots,234\}$, such that $\sum_k \delta t_k = T$. In  this  way we can write the evolution operator as $U(T,0)=\prod_k U(t_{k+1},t_k):=\prod_k U_k$, with each evolution block $U_k$ determined numerically. 
Experimentally, each $U_k$ is translated into a sequence of pulses and free system evolution (see figure~\ref{fig:simulation}). The time intervals $\delta t_k$ are not necessarily equal, and their choice take into account how fast the Hamiltonian changes during each block. Such an approach is necessary because there exists a trade-off between the minimization of the number of pulses and the fidelity obtained for the time evolution. Pulses' angles and the free evolution time interval were chosen as to minimize the (Hilbert-Schmidt) distance between $U_k$ and the implemented transformation. More details can be found in the Appendix B.

\begin{figure}[t!]
\begin{center}\includegraphics[width=\columnwidth]{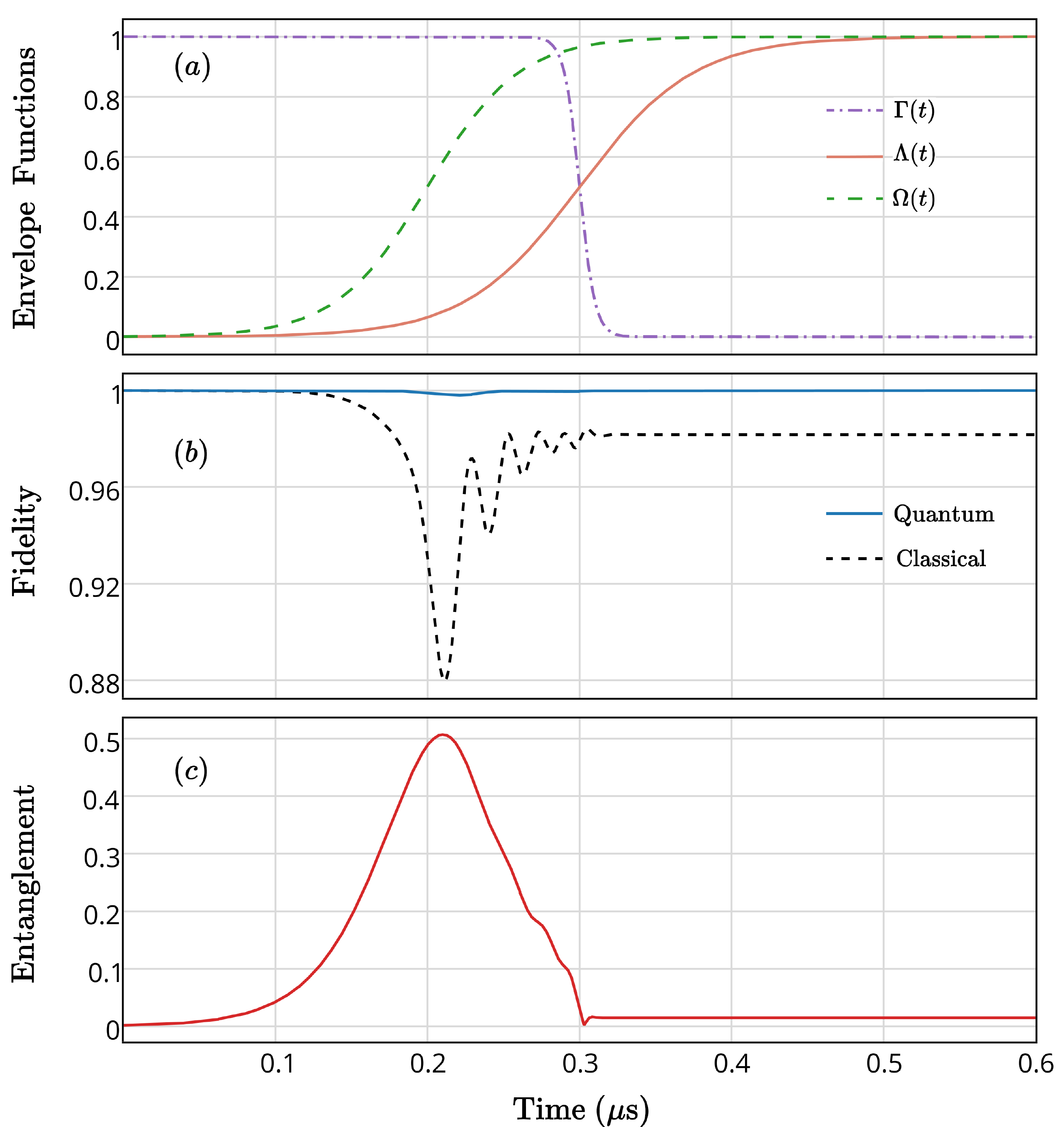}\end{center}
 \caption{\footnotesize \textbf{Ideal theoretical evolution (simulated protocol).} \textbf{(a) Scheduling.} Chosen profile for the envelope functions, where $\Gamma(t)=\left(1-\tanh[(t-0.3)/0.01]\right)/2$, $\Lambda(t)=\left(1+\tanh[(t-0.3)/0.075]\right)/2$ and $\Omega(t)=\left(1+\tanh[(t-0.2)/0.06]\right)/2$, with time in units of $\mu$s. The simulated protocol total time is fixed to $0.6\mu \text{s}$. \textbf{(b) Protocol fidelity.}  Quantum (solid line) and classical (dashed line) fidelities between the time-evolved state and the instantaneous ground state ($J_{12}<0$). The fidelity for the quantum protocol is degraded when the scheduling imposes fast Hamiltonian changes. For the classical protocol errors appear when some entanglement is expected in the ground state. \textbf{(c) Entanglement evolution.} The evolution of entanglement, as measured by its Negativity, in the time-evolved state. High amounts of entanglement are obtained when the contributions of $H_\text{easy}$ and $H_\text{Ising}$ are comparable. The results for the other instance ($J_{12}>0$) are roughly the same.} 
 \label{fig:setup}
\end{figure}

Here, we choose to experimentally search for the ground state of two instances of the Ising Hamiltonian~\eqref{ising} with parameters $h_1=62.5 \text{MHz}$ for the qubit encoded in the Hydrogen nuclear spin, and $h_2=15.625 \text{MHz}$ for the qubit encoded in the Carbon nuclear spin. The instances differ from each other only by the sign of spin-spin coupling $J_{12}$, which absolute value is chosen to be $53.75 \text{MHz}$.  Taking $|h_1|\neq |h_2|\neq|J_{12}|$ makes the ground state of $H_\text{Ising}$ unique and separable, as any possible degeneracy is lifted. For $H_\text{easy}$ we take $\Delta_1=28.125 \text{MHz}$ and $\Delta_2= 15.625 \text{MHz}$. Hence, any entanglement created during the time evolution is due to the adiabatic algorithm. With these choices, the total time for the simulated protocol was fixed in $0.6\mu s$  with  scheduling [see Fig.~\ref{fig:setup}(a)] selected in such a way that no level-crossing involving the ground state is present. In an ideal, noiseless and error-free realization of this protocol, fidelities between the time-evolved state and the instantaneous ground state would remain above 0.997 during the whole process, reaching  $\sim0.999$ at the end of the protocol (see Fig.~\ref{fig:setup}(b)). Moreover, in this scenario, the quantum annealer would always be better than the classical simulation described above, as may be seen in Fig.~\ref{fig:setup}(b). Also note, Fig.~\ref{fig:setup}(c), that when $\Gamma(t)H_\text{easy}$ and the spin-spin coupling Hamiltonian $\Omega(t)J_{12}\sigma_1^z\sigma_2^z$ are comparable, the instantaneous time-evolved state would exhibit a fair amount of entanglement. In addition to all the features highlighted above, the choice of parameters for our instances was such that the effective knob to adjust the noise level could lead to appreciable effects to the dynamics, allowing one to drawn conclusions from the results obtained.

As for the digitized implementation of the ideal protocol (Fig.~\ref{fig:setup}(a)), we found that our procedure to determine each evolution block $U_k$ gave us fidelities which were always in excess of $0.983$, demonstrating that a quantum computation performed using a banged-digital-analog implementation may not be necessarily compromised, even when considering a massive number of steps. With the experiment design fixed (scheduling and optimized $U_k$'s), our NMR implementation obeys the following structure: \emph{a)} we initialize the system in the (pseudo-pure) state $\ket{-_H}\ket{-_C}$; \emph{b)} switch on the field gradient; \emph{c)} apply a sequence of pulses leading to the evolution $\prod_{k}^n U_k$, up to a sequence with $(n\text{ mod } 3)=0$; \emph{d)} switch off the field gradient; \emph{e)} perform full-state tomography. This is repeated up to $n=234$ for a fixed field gradient. With the state snapshots we evaluate various quantities that characterize the quality of the computation and the entanglement generated. Afterwards we change the field gradient and the whole process is repeated (see Appendix B for more details about the state initialization and pulse sequence).

\begin{figure*}[t!]
\begin{center}
\includegraphics[scale=0.11]{Fig3a.pdf}\\
\vspace{0.2cm}
\hspace*{1cm}
\includegraphics[scale=0.25]{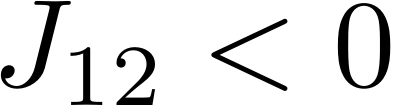}
\hspace{6.9cm}
\includegraphics[scale=0.25]{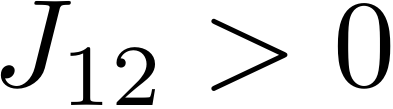}\\
\vspace{0.1cm}
\includegraphics[width=\columnwidth]{Fig3b.pdf}
\includegraphics[width=\columnwidth]{Fig3c.pdf}
\end{center}
 
 \caption{\footnotesize \textbf{Quantum annealing time evolution.} Experimental results obtained for both instances (left panels: $J_{12} <0$; right panels: $J_{12} >0$). Several figures of merit were evaluated for increasing values of applied magnetic field gradient: open squares show the fidelity between the experimentally measured state and the theoretical instantaneous ground state; the  success, solid disks, gives the fidelity as for the open squares, but only taking into account the diagonal part of the density matrices in the computational basis; open diamonds show the evolution of entanglement in in the experiment; and, finally,  blue crosses give information about the purity of the system. A general quality decrease is observed when the field gradient is increased.}
    \label{fig:results}
\end{figure*}

\section{Results and Discussion} Part of the results are shown in Fig. \ref{fig:results}. As expected, the fidelity (open squares) between the experimentally produced state and the ground state is near unity at the beginning of the experiment for all values of the gradient. As the protocol continues, fidelity decreases due to intrinsic errors and also due to the induced noise by the field gradient. Clearly, the greater the gradient, the worse the fidelity gets. Such a behavior is also observed when one looks at the fidelity of states considering only the population occupation of the computational basis (diagonal part of the state density matrix in the computational basis), what we called ``Success''. This figure of merit is pertinent for experiments where measurements can only be performed in the computational basis. Notice that towards the end of the protocol, both success and fidelity reach the same value. This happens because the off-diagonal terms of the experimental state die out due to the decoherence, and the ground state is diagonal in the computational basis.

\begin{figure*}[t!]
\begin{center}
\hspace*{1cm}
\includegraphics[scale=0.25]{JNeg.pdf}
\hspace{6.9cm}
\includegraphics[scale=0.25]{JPos.pdf}\\
\includegraphics[width=\columnwidth]{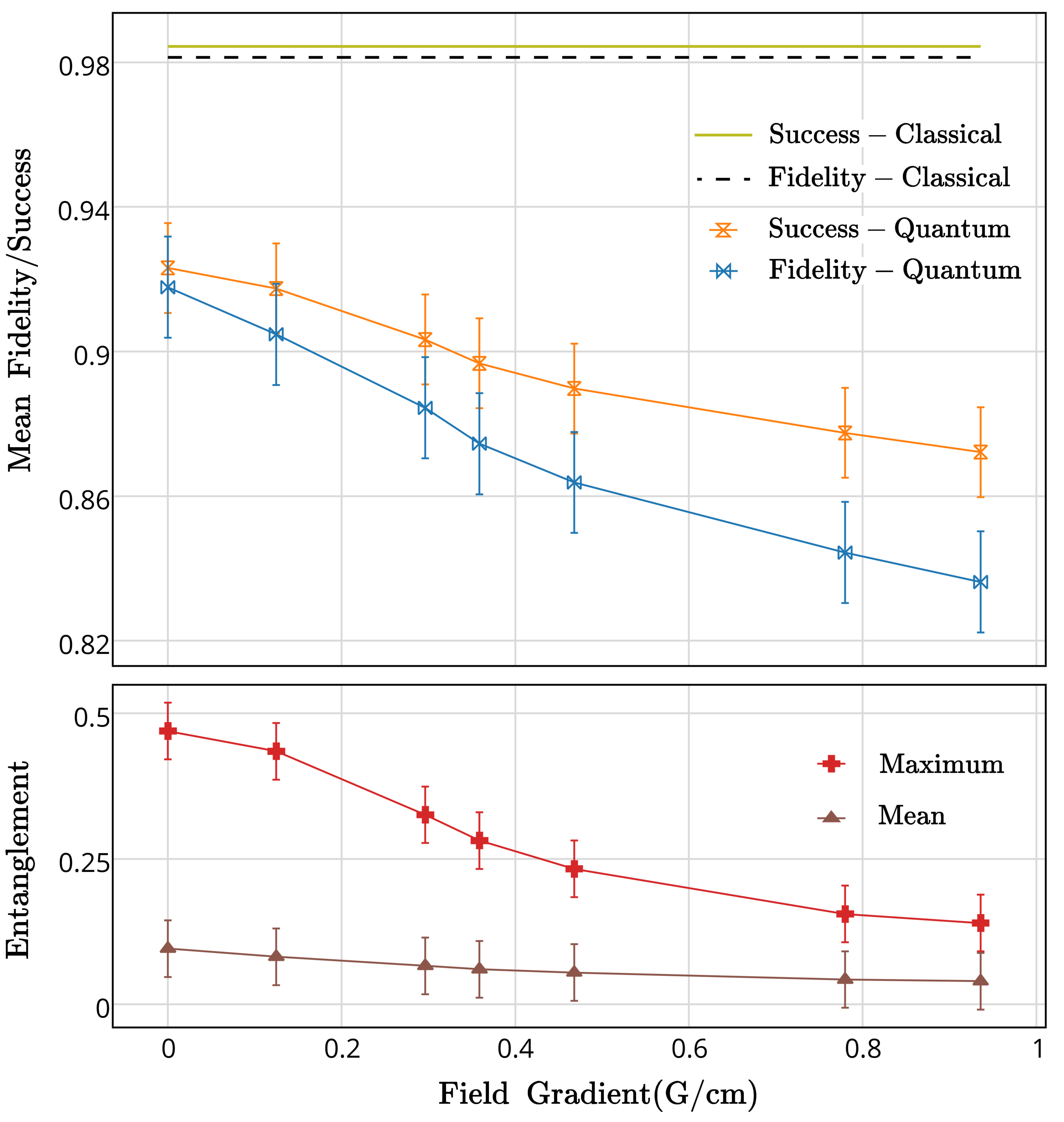}
\includegraphics[width=\columnwidth]{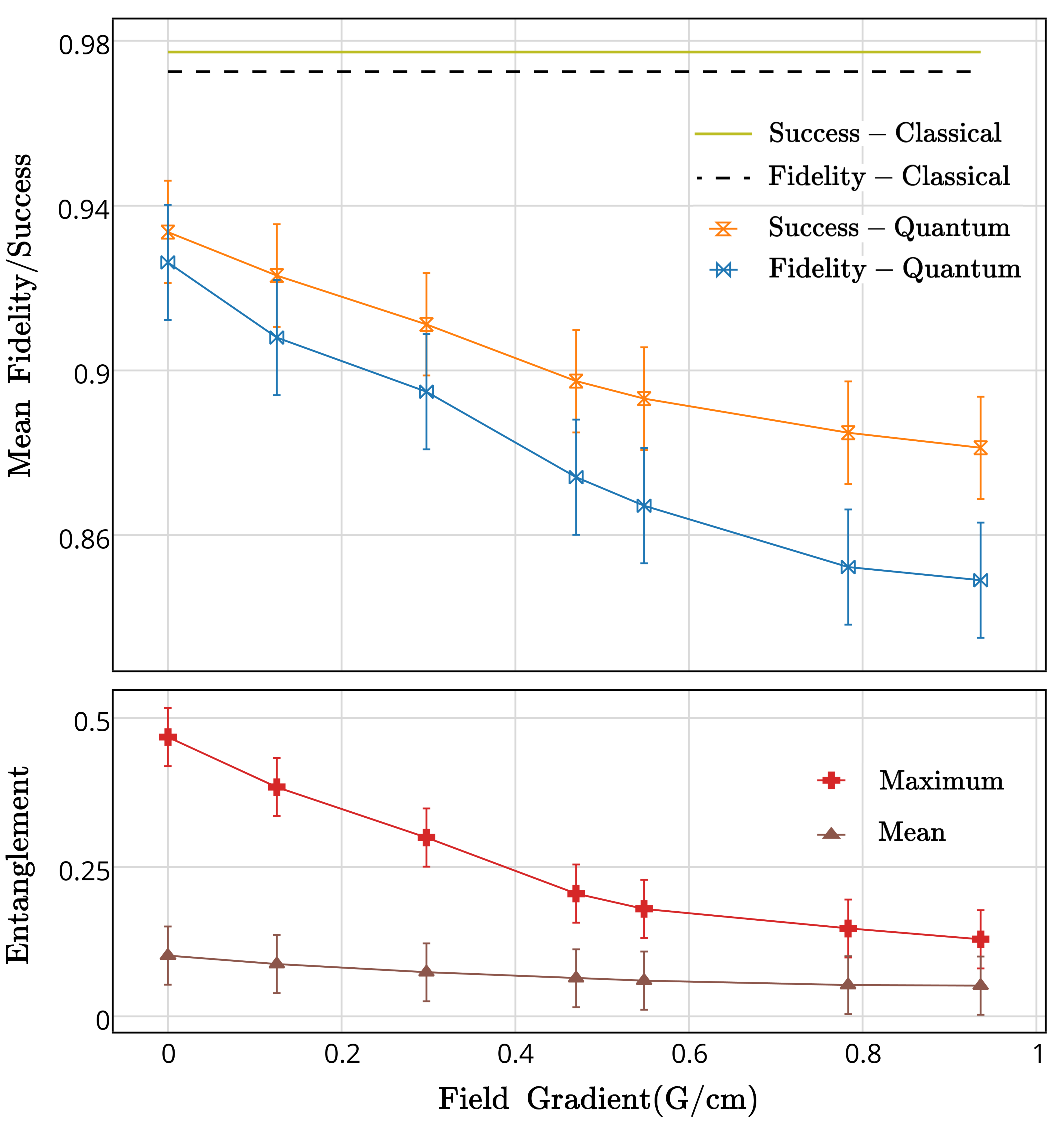}
\end{center}

 \caption{\footnotesize{\bf Computation Reliabilty.} Experimental results for mean (time-average) success and fidelity (top panel) and maximum and mean entanglement (bottom panel), as a function of the applied magnetic field gradient--recall that the noiseless and error-free realization of the digitized protocol gives mean fidelity (not shown in the panels) in excess of $0.993$ for both instances and hence above the classical algorithm. Classical algorithm is assumed to not suffer any kind of noise, as it is implemented in a classical computer. Left panels present the results for instance $J_{12} <0$, and the right panels for instance $J_{12} >0$.}
    \label{fig:AVG}
\end{figure*}

We can also observe how the amount of entanglement evolves with time, and its resilience to noise. The top panels of Fig.~\ref{fig:results}, where no gradient is present, is to be compared with Fig.\ref{fig:setup}(c)--recall that the zero field gradient implementation is still a noisy implementation of our protocol (see Appendix). As discussed before, when both the transversal and longitudinal components of the time-dependent Hamiltonian are present, a higher amount of entanglement is generated. As we increase the noise strength, i.e., the field gradient, the amount of entanglement created during the protocol greatly reduces.  These two facts above suggest a correlation between the amount of entanglement generated during the quantum annealing protocol and the overall quality of the process. To make this correlation clearer, in Fig.~\ref{fig:AVG} we plot the time-average fidelity and success for the quantum annealer as a function of the field gradient, and also the time-averaged entanglement and the maximum achieved entanglement as a function of the field gradient. The curves are monotonically related to each other.

Naturally, the inference of such a correlation between the entanglement generated and the quality of the process poses perhaps a more important question regarding quantum computation: would the presence of entanglement be a signature of a better than classical computation? To address this question, we also plot in Fig.~\ref{fig:AVG} the  time-average fidelity and success for the classical simulation. As the ground state of $H(t)$ is entangled at some times, and the states produced by the classical algorithm are always separable, the time averaged fidelity and success can never be one. Nevertheless, as the classical simulation does not suffer from noise, its quality is the same for all values of the gradient, registering fidelity values of $0.981$ and $0.972$ for instance $J_{12}<0$ and $J_{12}>0$, respectively. Since the noiseless and error-free realization of the digitized protocol gives fidelity always above $0.983$, which leads to mean fidelity values in excess of $0.993$ for both instances, one finds that the ideal implementation of the quantum computation surpasses its classical analog. However, with the quality degradation of the quantum algorithm in noisy implementations, it could be expected that for a given level of noise the quantum fidelity and success could be below the classical counterparts. Surprisingly, this crossing happens despite the fact that considerable amounts of entanglement are still generated by the quantum annealer, as one can observe in Fig.~\ref{fig:AVG}. This shows that producing entanglement in a quantum annealing process does not necessarily mean that the quantum computation is more reliable than any classical simulation.

\section{Conclusions} Our results, obtained from an NMR quantum annealer, provide clear evidences that entanglement should not be considered the figure of merit to assert that a quantum annealing computation would be more reliable than any classical computation, even though a correlation between high amounts of entanglement and better quantum computation seems to exist. In addition, as for the context of digital-analog adiabatic quantum computing, our results also reinforce that such an approach should be indeed considered as a viable and promising implementation of continuous time evolutions. 
The combination of a long analog part with a fault-tolerant sequence of digital gates might pave  the way to bring the promise of a quantum computer to a more tangible reality.

\section {\bf Acknowledgements}
We would like to thank Enrique Solano for various important remarks to our results. 
The authors are supported by the Instituto Nacional de Ci\^encia e Tecnologia - Informa\c{c}\~ao Qu\^antica (INCT-IQ), and by the Brazilian funding agencies FAPERJ, and CNPq. Author contributions: F.~B.~and F.~d-M.~equally contributed to this project.

\section{ Appendix A: NMR Setup}

\subsection{Radio-Frequency Pulses}

To manipulate the spins during an NMR experiment we apply radio-frequency pulses of an oscillating magnetic field $\mathbf{B}_{1}$. This field is applied in the $xy$ plane and has a much smaller intensity than the field $\mathbf{B}_{0}$. To the Hamiltonian~\eqref{eq:Hnmr} we thus add the time-dependent radio-frequency pulse Hamiltonian, namely:
\begin{equation}
H_{\rm rf}(t)=\frac{\hbar \omega_1(t)}{2}\big\{ \cos(\omega_\text{rf} t+\phi) \sigma_{x}+\sin(\omega_\text{rf}t+\phi)\sigma_{y}\big\} .
\label{Hrf}
\end{equation}
Where $\omega_1 = -\gamma |B_1|$  controls the pulse shape, with $\gamma$ representing the gyromagnetic ratio for the nuclear spin. Tuning $\omega_\text{rf}$ close to $\omega_H$ or $\omega_C$ allows one to select the nuclei to act upon. As the nuclear Larmor frequencies are far from each other, a quasi-squared short pulse could be used. The pulse amplitude, $\omega_1$, in our experiment was set such that a $\pi/2$ rotation in the Hydrogen nuclear spin is performed in $33.5\mu\text{s}$. In the rotating frame of each nuclei, this quasi-squared pulse translates into a spin rotation $R(\phi,t)=\exp[\ii \omega_1 t (\cos(\phi)\sigma_x +\sin(\phi)\sigma_y)/2]$,  with the phase $\phi$ setting the spin precession axis.\\

\subsection{Relaxation and Decoherence}

As usual in NMR experiments, ours was performed at room temperature ($\sim 25{\rm ^oC}$). Under such an ordinary condition, the natural nuclei thermalization process leads to observed $T_1$ relaxation and $T_2$ decoherence time scales given by   $(T_1^{(H)}=7.4{\rm s},~T_2^{(H)}=0.245{\rm s})$ and $(T_1^{(C)}=11.3{\rm s},~T_2^{(C)}=0.157{\rm s})$.

Upon adding a magnetic field gradient along the $z$ direction (i.e. along the static field $\mathbf{B}_0$), one is capable of increasing the decoherence process experienced by both nuclei on demand. Indeed, under such a condition the nuclei Larmor frequencies become $(z-)$position dependent, imposing different spin precessions and therefore enhancing the dephasing observed in the system dynamics (\ref{eq:Hnmr}). This dephasing channel can be seen more quantitative when considering the case of a sample of size $L$, comprised of one species of nuclear spin, in the presence of an applied magnetic field given by $\mathbf{B}(z)=(B_0+z\frac{\partial B}{\partial z})\hat{z}$. Under this condition, by means of techniques available in an NMR experiment, one only has access to the time-evolved system density matrix as the mixture state $\rho_{{\rm NMR}}(t)=(1/L)\int_0^Ldzu_z\rho_0(t) u_z^\dagger$, where $\rho_0(t)$ represents the evolved density matrix for the ideal case of non-field gradient applied, and  $u_z=\exp{\left(i t\gamma z\frac{\partial B}{\partial z}\sigma^z\right) }$. Accordingly, one finds that the off-diagonal element of $\rho_{{\rm NMR}}(t)$ -represented in the $\sigma^z$ basis- will decay in time as $\exp(-\log( t\gamma L\frac{\partial B}{\partial z}))$, which explicitly illustrates that the lack of spatial resolution in NMR can lead to the observation of a dephasing process when a magnetic field gradient is present. The gradient was characterized by a standard diffusion experiment known as Pulsed Gradient Spin Echo~\cite{Stejskal} and was always set to be below $1\rm G/cm$.\\

\begin{figure*}[t!]
	\includegraphics[width=\linewidth]{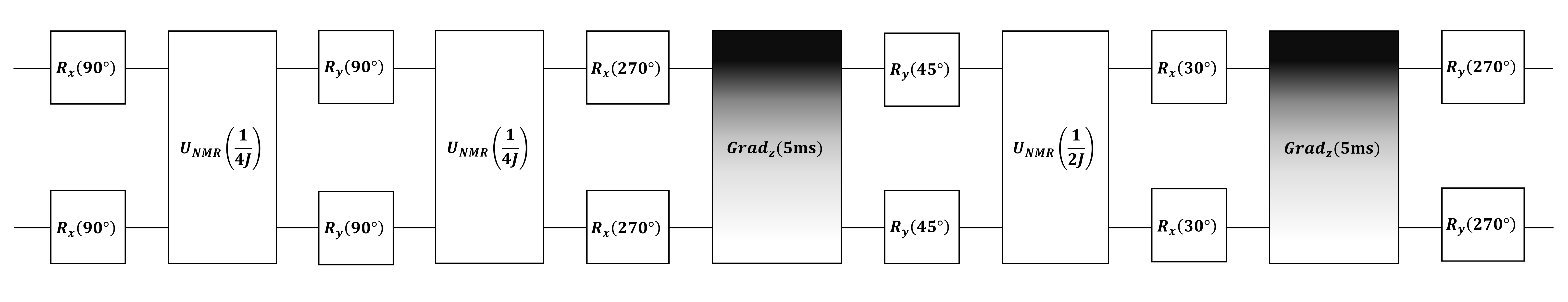}
	\caption{Set of operations needed to be applied at each qubit (lines) to prepare
		the pseudo-pure state $\rho=\frac{(1-\epsilon)}{4}\idty+\epsilon\left|--\right\rangle \left\langle --\right|$. In the picture, 	$R_{\alpha}(\theta)$ stands for rotations around the axis $\alpha$
		by an angle $\theta$, $U_\text{NMR}(t)$ represents the free evolution for a time $t$, and $Grad_{z}(t)$ represents the application of a strong field gradient applied along the $z$ direction with duration $t$.\label{fig:pseudoPure}}
\end{figure*}

\subsection{Measurements}

The observable quantity in NMR is the temporal change of the sample magnetization. Roughly speaking, the nuclei spin precession of the molecules in the sample produce an oscillating magnetic field which is attenuated by the thermalization processes discussed above. This measurement process is known as \emph{Free Induction Decay} (FID), and more details can be found in~\cite{IvanLivro}. In our setup the signal is measured by coils parallel to the sample  -- the same coils which are used to produce the radio frequency pulses. In this way only the $x$ and $y$ components of the magnetization are measured, and the transversal relaxation (decoherence) is the main factor in the signal decay. Each coil has a local oscillator in the resonance frequency of each nuclei (that is how the rotating frame is established), and thus one coil allows for determining the $\<\sigma_x\>$ and $\<\sigma_y\>$ components of the Hydrogen nuclear spin, while the other measures the same components of the Carbon nuclear spin.  The measurement of these components is not enough to perform full state tomography. To accomplish a tomographic process, rotations are applied on the sample before the measurement. The rotations that need to be applied in the case of two qubits can be found in~\cite{PhysRevA.69.052302}.\\

\section{Appendix B: The experiment}

As described in the main text, in order to probe the state of the system at a given time $t$  for a fixed noise level, our experiment abides by the following  algorithm: \emph{a)} Prepare the initial state $\ket{--}$, which is the ground state of the easy Hamiltonian; \emph{b)} Switch on the magnetic field gradient; \emph{c)} Apply a sequence of pulses to simulate the dynamics given by~\eqref{eq:timeH} up to time $t$; \emph{d)} Switch off the magnetic field gradient; \emph{e)} Perform tomographic measurements. These steps are repeated until all the measurements to complete the state tomography are performed. We then repeat the whole algorithm for the next time step.

Below we describe in more details points \emph{a)} and \emph{d)}, as the others were already delineated above. Besides, we show further results not presented in the main text, and discuss the assessment of experimental errors. \\

\subsection{State initialization}

As mentioned above, the experiments are performed at room temperature. Each molecule, before the experiment, is thus in the thermal equilibrium state. In NMR one cannot execute projective measurements, and the only available operations one can conduct on the nuclear spins are rotations and a gradient-induced dephasing. With these constraints one cannot prepare a pure state. We thus resort to a \emph{pseudo-pure} state~\cite{Gershenfeld350}. We use the sequence of pulses shown in Fig.~\ref{fig:pseudoPure} to prepare the state $\rho=\frac{(1-\epsilon)}{4}\idty+\epsilon\left|--\right\rangle \left\langle --\right|$. At room temperature we have $\epsilon\sim10^{-5}$. It is worth emphasizing that the part proportional to the identity does not interfere or gets measured in a NMR experiment -- the available operations have no effect on the identity; moreover, only traceless observables are measured, and thus the identity does not contribute to any signal.\\

\subsection{Pulse sequence}

In order to simulate the evolution of the system determined by the Hamiltonian~\eqref{eq:timeH} we used  a sequence of radio-frequency pulses and free evolutions. For that, the total duration of the desired simulated dynamics, $0.6\mu s$, was divided in 235 time steps. To each time step corresponds an evolution operator $U_k$, with $k\in\{0,\ldots,234\}$, which we evaluated theoretically. To determine the sequence of pulses we numerically minimized the square of the Hilbert-Schmidt distance between $U_k$ (the exact unitary) and $U_k^{\text{exp}}$ (the implemented unitary), i.e., we minimized $||U_k-U_k^{\text{exp}}||_2^2 = 2\{4-\Re[\tr(U_k^{\text{exp}}.U_k^\dagger)]\}$. To do that we parametrized $U_k^{\text{exp}}$ as \begin{widetext}
\begin{equation*}
U_k^\text{exp} = \Big(R_x(\theta_1^k)\otimes R_x(\theta_2^k)\Big).\Big(e^{-\frac{\ii}{\hbar}H_\text{NMR}^\prime\Delta\!t_k}\Big).\Big(R_z(\theta_3^k)\otimes R_z(\theta_4^k)\Big).\Big(R_x(\theta_1^k)\otimes R_x(\theta_2^k)\Big),
\end{equation*}\end{widetext} where $H_\text{NMR}^\prime = 2\pi\hbar J \sigma^z_H \sigma^z_C/4$ is the natural NMR Hamiltonian in the rotating frame.

\begin{figure}[t]
        \includegraphics[width=\linewidth]{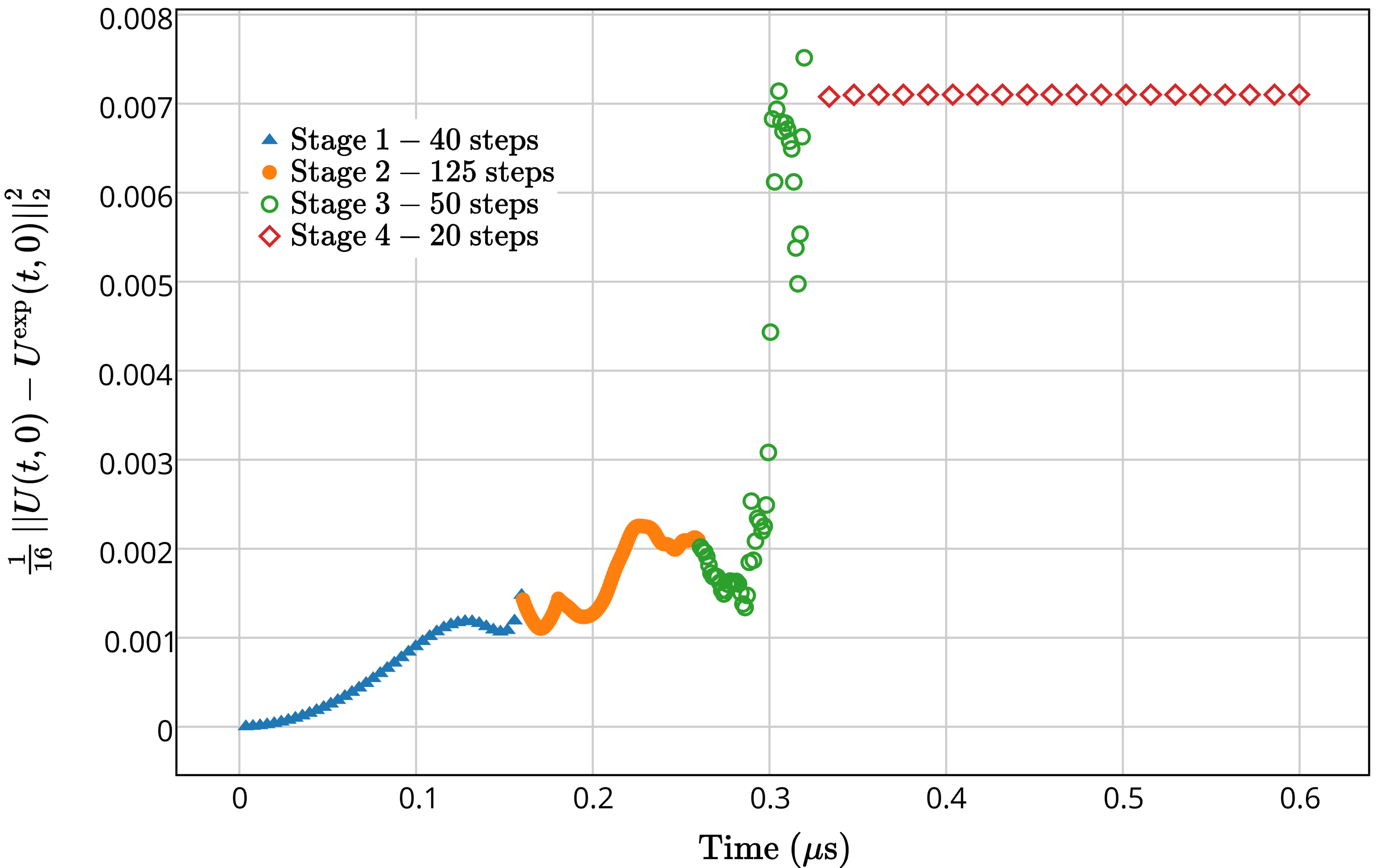}
	\caption{\label{fig:HS} Normalized Hilbert-Schmidt distance between the exact and the approximated evolution operator as a function of simulated time. Each one of the four stages has its specific number of steps determined in order to minimize the distance.}
\end{figure}

The errors incurred due to that procedure can be minimized by adjusting the number of steps around the time intervals where the Hamiltonian changes the most. We thus divided the whole evolution in four stages, each one being divided by a specific uniform time step.  See Fig.~\ref{fig:HS}.

\begin{figure*}[t]
	\includegraphics[width=\linewidth]{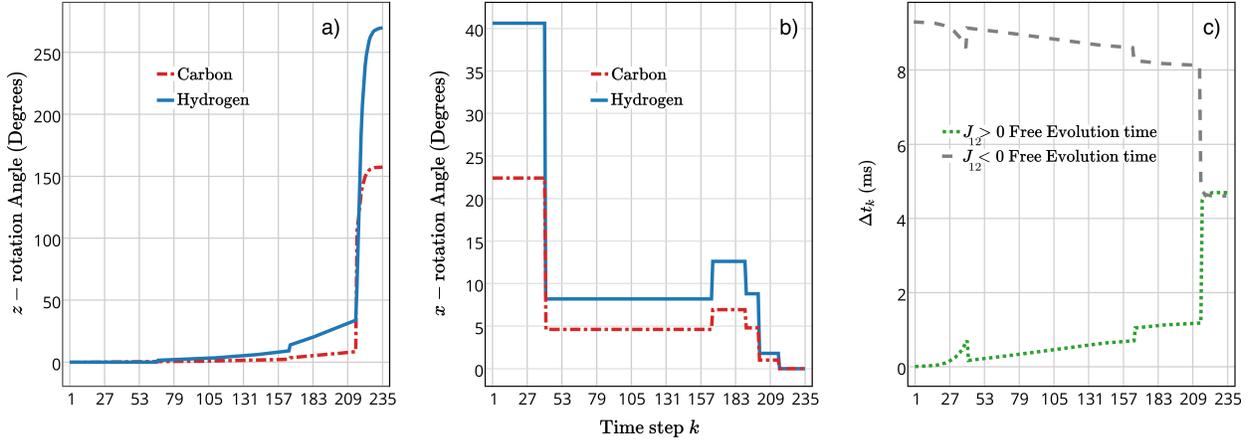}
	\caption{\label{fig:angles} Values used in the experiment for the evolution operator $U_k^\text{exp}$ as a function of the time step $k$. a) and b) Angles for the rotation pulses in $z$ and $x$ directions, respectively. c) Free evolution times $\Delta\!t _k$. The digital part for both instances is identical, with the same angles for the pulses. The difference between the simulation of the instances is completely encoded in the analog part of the protocol.}
\end{figure*}

This minimization generated some very small angles, whose experimental implementation could compromise the overall fidelity expected, besides of being below the experimental precision. Because of that, upon the top of the Trotterization procedure, we neglected any rotation with angles below $0.1\rm^o$. The Hilbert-Schmidt distance between the exact and the approximated evolution operator under such an approximation is shown in Fig.\ref{fig:HS}.  The values  used in the experiments can be read from the plots in Fig.\ref{fig:angles}. 

Lastly, as we cannot perform pulses in the $z$-axis, we used the mathematical identity $R_z(\theta)=R_x(\pi/2) R_y(\theta) R_x(-\pi/2)$. Therefore, each pulse along the $z$ direction was turned into a sequence of three pulses in the $xy$ plane. All in all, about 2000 pulses, in a $300\rm ms$ real-time evolution, were necessary to simulate the whole annealing protocol.\\

\subsection{Experimental Errors}

We assessed the errors present in the combined experimental procedure of preparing and measuring the system state by repeating it several times. By doing this, we found that the reconstructed density matrix elements were Gaussian distributed with a relative standard deviation of $0.1\%$. It is worth mentioning that such a procedure is performed just once during the evolution and hence its associated error will not scale as the number of time steps increase. 

As for the errors due to the fault pulses, which clearly scale up as the number of time steps increases, one should expect in an NMR experiment a typical error of few degrees for their angles. Indeed, our estimations revealed that standard deviations of $1\rm^o$ for rotations in the $x$ and $y$ directions and of $10\rm ns$ for the free evolution duration could correctly describe our findings. The error bars shown in Figs.\ref{fig:results} and \ref{fig:AVG} are evaluated correspondingly for such source of errors.


\section*{References}

\begin{thebibliography}{10}
	
	\bibitem{Farhi20042001}
	Edward Farhi, Jeffrey Goldstone, Sam Gutmann, Joshua Lapan, Andrew Lundgren,
	and Daniel Preda.
	\newblock A quantum adiabatic evolution algorithm applied to random instances
	of an np-complete problem.
	\newblock {\em Science}, 292(5516):472--475, 2001.
	
	\bibitem{Johnson:2011gd}
	M~W Johnson, M~H~S Amin, S~Gildert, T~Lanting, F~Hamze, N~Dickson, R~Harris,
	A~J Berkley, J~Johansson, P~Bunyk, E~M Chapple, C~Enderud, J~P Hilton,
	K~Karimi, E~Ladizinsky, N~Ladizinsky, T~Oh, I~Perminov, C~Rich, M~C Thom,
	E~Tolkacheva, C~J~S Truncik, S~Uchaikin, J~Wang, B~Wilson, and G~Rose.
	\newblock {Quantum annealing with manufactured spins}.
	\newblock {\em Nature}, 473(7346):194--198, May 2011.
	
	\bibitem{Barends2016}
	R~Barends, A~Shabani, L~Lamata, J~Kelly, A~Mezzacapo, U~Las Heras, R~Babbush,
	G~Fowler, B~Campbell, Yu~Chen, Z~Chen, B~Chiaro, A~Dunsworth, E~Jeffrey,
	E~Lucero, A~Megrant, J~Y Mutus, M~Neeley, C~Neill, P~J J~O Malley,
	C~Quintana, P~Roushan, A~Vainsencher, J~Wenner, T~C White, E~Solano, H~Neven,
	and John~M Martinis.
	\newblock {Digitized adiabatic quantum computing with a superconducting
		circuit}.
	\newblock {\em Nature}, 534(7606):222--226, 2016.
	
	\bibitem{Boixo:1ha}
	Sergio Boixo, Tameem Albash, Federico~M Spedalieri, Nicholas Chancellor, and
	Daniel~A Lidar.
	\newblock {Experimental signature of programmable quantum annealing}.
	\newblock {\em Nature Communications}, 4:1--8, 1.
	
	\bibitem{Dickson:1bv}
	N~G Dickson, M~W Johnson, M~H Amin, R~Harris, F~Altomare, A~J Berkley, P~Bunyk,
	J~Cai, E~M Chapple, P~Chavez, F~Cioata, T~Cirip, P~deBuen, M~Drew-Brook,
	C~Enderud, S~Gildert, F~Hamze, J~P Hilton, E~Hoskinson, K~Karimi,
	E~Ladizinsky, N~Ladizinsky, T~Lanting, T~Mahon, R~Neufeld, T~Oh, I~Perminov,
	C~Petroff, A~Przybysz, C~Rich, P~Spear, A~Tcaciuc, M~C Thom, E~Tolkacheva,
	S~Uchaikin, J~Wang, A~B Wilson, Z~Merali, and G~Rose.
	\newblock {Thermally assisted quantum annealing of a 16-qubit problem}.
	\newblock {\em Nature Communications}, 4:1903--6, 1.
	
	\bibitem{Boixo:2014cg}
	Sergio Boixo, Troels~F Ronnow, Sergei~V Isakov, Zhihui Wang, David Wecker,
	Daniel~A Lidar, John~M Martinis, and Matthias Troyer.
	\newblock {Evidence for quantum annealing with more than one hundred qubits}.
	\newblock {\em Nature Physics}, 10(3):218--224, February 2014.
	
	\bibitem{PhysRevA.91.042314}
	Tameem Albash, Walter Vinci, Anurag Mishra, Paul~A. Warburton, and Daniel~A.
	Lidar.
	\newblock Consistency tests of classical and quantum models for a quantum
	annealer.
	\newblock {\em Phys. Rev. A}, 91:042314, Apr 2015.
	
	\bibitem{10.3389/fphy.2014.00052}
	John~A Smolin and Graeme Smith.
	\newblock Classical signature of quantum annealing.
	\newblock {\em Frontiers in Physics}, 2(52), 2014.
	
	\bibitem{2014arXiv1401.7087S}
	S.~W. {Shin}, G.~{Smith}, J.~A. {Smolin}, and U.~{Vazirani}.
	\newblock {How ''Quantum'' is the D-Wave Machine?}
	\newblock {\em arXiv:1401.7087}, January 2014.
	
	\bibitem{2014arXiv1404.6499S}
	S.~W. {Shin}, G.~{Smith}, J.~A. {Smolin}, and U.~{Vazirani}.
	\newblock {Comment on ''Distinguishing Classical and Quantum Models for the
		D-Wave Device''}.
	\newblock {\em arXiv:1404.6499}, April 2014.
	
	\bibitem{Ronnow:2014fd}
	Troels~F Ronnow, Zhihui Wang, Joshua Job, Sergio Boixo, Sergei~V Isakov, David
	Wecker, John~M Martinis, Daniel~A Lidar, and Matthias Troyer.
	\newblock {Defining and detecting quantum speedup}.
	\newblock {\em Science}, 345(6195):420--424, 2014.
	
	\bibitem{PhysRevX.4.021041}
	T.~Lanting, A.~J. Przybysz, A.~Yu. Smirnov, F.~M. Spedalieri, M.~H. Amin, A.~J.
	Berkley, R.~Harris, F.~Altomare, S.~Boixo, P.~Bunyk, N.~Dickson, C.~Enderud,
	J.~P. Hilton, E.~Hoskinson, M.~W. Johnson, E.~Ladizinsky, N.~Ladizinsky,
	R.~Neufeld, T.~Oh, I.~Perminov, C.~Rich, M.~C. Thom, E.~Tolkacheva,
	S.~Uchaikin, A.~B. Wilson, and G.~Rose.
	\newblock Entanglement in a quantum annealing processor.
	\newblock {\em Phys. Rev. X}, 4:021041, May 2014.
	
	\bibitem{Arrazola}
	L.~Lamata I.~Arrazola, J. S.~Pedernales and E.~Solano.
	\newblock Digital-analog quantum simulation of spin models in trapped ions.
	\newblock {\em Sci. Rep.}, 6:30534, 2016.
	
	\bibitem{Lamata}
	Lucas Lamata.
	\newblock Digital-analog quantum simulation of generalized dicke models with
	superconducting circuits.
	\newblock 2016.
	
	\bibitem{Lloyd1073}
	Seth Lloyd.
	\newblock Universal quantum simulators.
	\newblock {\em Science}, 273(5278):1073--1078, 1996.
	
	\bibitem{Barahona:1982gj}
	F~Barahona.
	\newblock {On the computational complexity of Ising spin glass models}.
	\newblock {\em Journal of Physics A: Mathematical and General},
	15(10):3241--3253, October 1982.
	
	\bibitem{messiah}
	A.~Messiah.
	\newblock {\em Quantum Mechanics}.
	\newblock Dover Publications, Mineola, 1999.
	
	\bibitem{matthias}
	Matthias Steffen, Wim van Dam, Tad Hogg, Greg Breyta, and Isaac Chuang.
	\newblock Experimental implementation of an adiabatic quantum optimization
	algorithm.
	\newblock {\em Phys. Rev. Lett.}, 90:067903, Feb 2003.
	
	\bibitem{NMRreview}
	R.~Laflamme, E.~Knill, D.~G. Cory, E.~M. Fortunato, T.~Havel, C.~Miquel,
	R.~Martinez, C.~Negrevergne, G.~Ortiz, M.A. Pravia, Y.~Sharf, S.~Sinha,
	R.~Somma, and L.~Viola.
	\newblock Introduction to {NMR} quantum information processing.
	\newblock {\em Los Alamos Science}, 27:226 -- 259, 2001.
	
	\bibitem{IvanLivro}
	I.~Oliveira, R.~Sarthour Jr., T.~Bonagamba, E.~Azevedo, and J.~C.~C. Freitas.
	\newblock {\em NMR quantum information processing}.
	\newblock Elsevier, 2007.
	
	\bibitem{Stejskal}
	E.~O. Stejskal and J.~E. Tanner.
	\newblock Spin diffusion measurements: Spin echoes in the presence of a
	time-dependent field gradient.
	\newblock {\em The Journal of Chemical Physics}, 42(1):288--292, 1965.
	
	\bibitem{PhysRevA.69.052302}
	Garett~M. Leskowitz and Leonard~J. Mueller.
	\newblock State interrogation in nuclear magnetic resonance quantum-information
	processing.
	\newblock {\em Phys. Rev. A}, 69:052302, May 2004.
	
	\bibitem{Gershenfeld350}
	Neil~A. Gershenfeld and Isaac~L. Chuang.
	\newblock Bulk spin-resonance quantum computation.
	\newblock {\em Science}, 275(5298):350--356, 1997.
	
\end{thebibliography}
\end{document}